\documentclass[twocolumn,prl,showpacs,preprintnumbers,amsmath,amssymb]{revtex4}

\usepackage{graphicx}

\begin{document}


\title{Universal \textit{versus} Material-Dependent Two-Gap Behaviors in the High-$T_c$ Cuprates: Angle-Resolved Photoemission Study of La$_{2-x}$Sr$_x$CuO$_4$}

\author{T. Yoshida$^1$, M. Hashimoto$^1$, S. Ideta$^1$, A. Fujimori$^1$, K. Tanaka$^2$, N. Mannella$^2$,
Z. Hussain$^3$, Z.-X. Shen$^2$, M. Kubota$^4$, K. Ono$^4$, Seiki
Komiya$^5$, Yoichi Ando$^6$, H. Eisaki$^7$, S. Uchida$^1$}
\affiliation{$^1$Department of Physics, University of Tokyo,
Bunkyo-ku, Tokyo 113-0033, Japan}\affiliation{$^2$Department of
Applied Physics and Stanford Synchrotron Radiation Laboratory,
Stanford University, Stanford, CA94305} \affiliation{$^3$Advanced
Light Source, Lawrence Berkeley National Lab, Berkeley, CA 94720}
\affiliation{$^4$Photon Factory, Institute of Materials Structure
Science, KEK, Tsukuba, Ibaraki 305-0801, Japan}
\affiliation{$^5$Central Research Institute of Electric Power
Industry, Komae, Tokyo 201-8511, Japan} \affiliation{$^6$Institute
of Scientific and Industrial Research, Osaka University, Ibaraki,
Osaka 567-0047, Japan} \affiliation{$^7$National Institute of
Advanced Industrial Science and Technology, Tsukuba 305-8568,
Japan}
\date{\today}

\begin{abstract}
We have investigated the doping and temperature dependences of the
pseudogap/superconducting gap in the single-layer cuprate
La$_{2-x}$Sr$_x$CuO$_4$ by angle-resolved photoemission
spectroscopy. The results clearly exhibit two distinct energy and
temperature scales, namely, the gap around ($\pi$,0) of magnitude
$\Delta^*$ and the gap around the node characterized by the
$d$-wave order parameter $\Delta_0$, like the double-layer cuprate
Bi2212. In comparison with Bi2212 having higher $T_c$'s,
$\Delta_0$ is smaller, while $\Delta^*$ and $T^*$ are similar.
This result suggests that $\Delta^*$ and $T^*$ are approximately
material-independent properties of a single CuO$_2$ plane, in
contrast the material-dependent $\Delta_0$, representing the
pairing strength.
\end{abstract}

\pacs{74.25.Jb, 71.18.+y, 74.72.Dn, 79.60.-i}

\maketitle

 One of the central issues in the studies of high-$T_c$ cuprates
is whether the pseudogap is related to the superconductivity or a
distinct phenomenon from superconductivity. In the former
scenario, a possible origin of the pseudogap is preformed Cooper
pairs lacking phase coherence \cite{Kivelson}. In the latter
scenario, the pseudogap is due to a competing order such as spin
density wave, charge density wave, $d$-density wave \cite{DDW},
etc. It has been well known that the pseudogap in the antinodal
$\sim$ ($\pi$,0) region increases with underdoping as observed by
angle-resolved photoemission spectroscopy (ARPES) \cite{Campuzano}
and tunneling spectroscopy \cite{Miyakawa}. However, the energy
gap measured by Andreev reflection \cite{Deustcher}, penetration
depth \cite{Panagopoulos}, and Raman experiments in
B$_{2g}$-geometry \cite{Opel,Tacon}, which is more directly
associated with superconductivity, exhibits opposite trend, that
is, the gap decreases with underdoping, suggesting a different
origin of the superconducting gap from the antinodal gap.

A recent ARPES study of deeply underdoped
Bi$_2$Sr$_2$Ca$_{1-x}$Y$_x$Cu$_2$O$_8$ (Bi2212) has revealed the
presence of two distinct energy gaps between the nodal and
anti-nodal region \cite{Tanaka,Lee}. A similar two-gap behavior
has been observed in optimally doped single-layer cuprate
Bi$_2$Sr$_2$CuO$_{6+\delta}$ (Bi2201)
\cite{Kondo,HashimotoDisorder} and La$_{2-x}$Sr$_x$CuO$_4$ (LSCO)
\cite{Terashima}. Also, a temperature-dependent angle-integrated
photoemission study of LSCO has indicated two distinct gap energy
scales \cite{Hashimoto}. On the other hand, attempts have been
made to understand the pseudogap within a single $d$-wave energy
gap \cite{NormanArc,Valla,Kanigel,Meng}. Valla \textit{et al.}
\cite{Valla} have shown that La$_{2-x}$Ba$_x$CuO$_4$ with $x$=1/8,
where superconductivity is suppressed due to stripe formation, has
a gap of simple $d_{x^2-y^2}$ symmetry without signature of two
gap energy scales. From the measurement of Fermi arc length,
Kanigel \textit{et al.} \cite{Kanigel} has proposed that the $T$
=0 ground state of the pseudogap state is a nodal liquid which has
a single $d_{x^2-y^2}$ gap. In such a single gap picture,
preformed Cooper pairs are the most likely origin of the
pseudogap.

Since the doping and temperature dependences of the energy gap
would reveal the entangled two-gap behavior, we have investigated
the energy gap of lightly- to optimally-doped LSCO by ARPES as a
function of doping and temperature. In the present work, the
momentum dependence of the gap clearly exhibits two-gap behavior
as in the case of heavily underdoped Bi2212: the pseudogap
$\Delta^*$ in the antinodal region and the $d$-wave like gap
$\Delta_0$ around the node. Furthermore, from comparison of the
present results with those on Bi2212, we have found that the
magnitude of the $\Delta^*$ and the pseudogap temperature $T^*$ is
not appreciably material-dependent, suggesting that the pseudogap
is properties of a single CuO$_2$ plane. On the other hand, the
magnitude of the $\Delta_{0}$, which is proportional to the
superconducting gap, is strongly material-dependent (CuO$_2$ layer
number-dependent) like $T_c$.

\begin{figure}
\includegraphics[width=9cm]{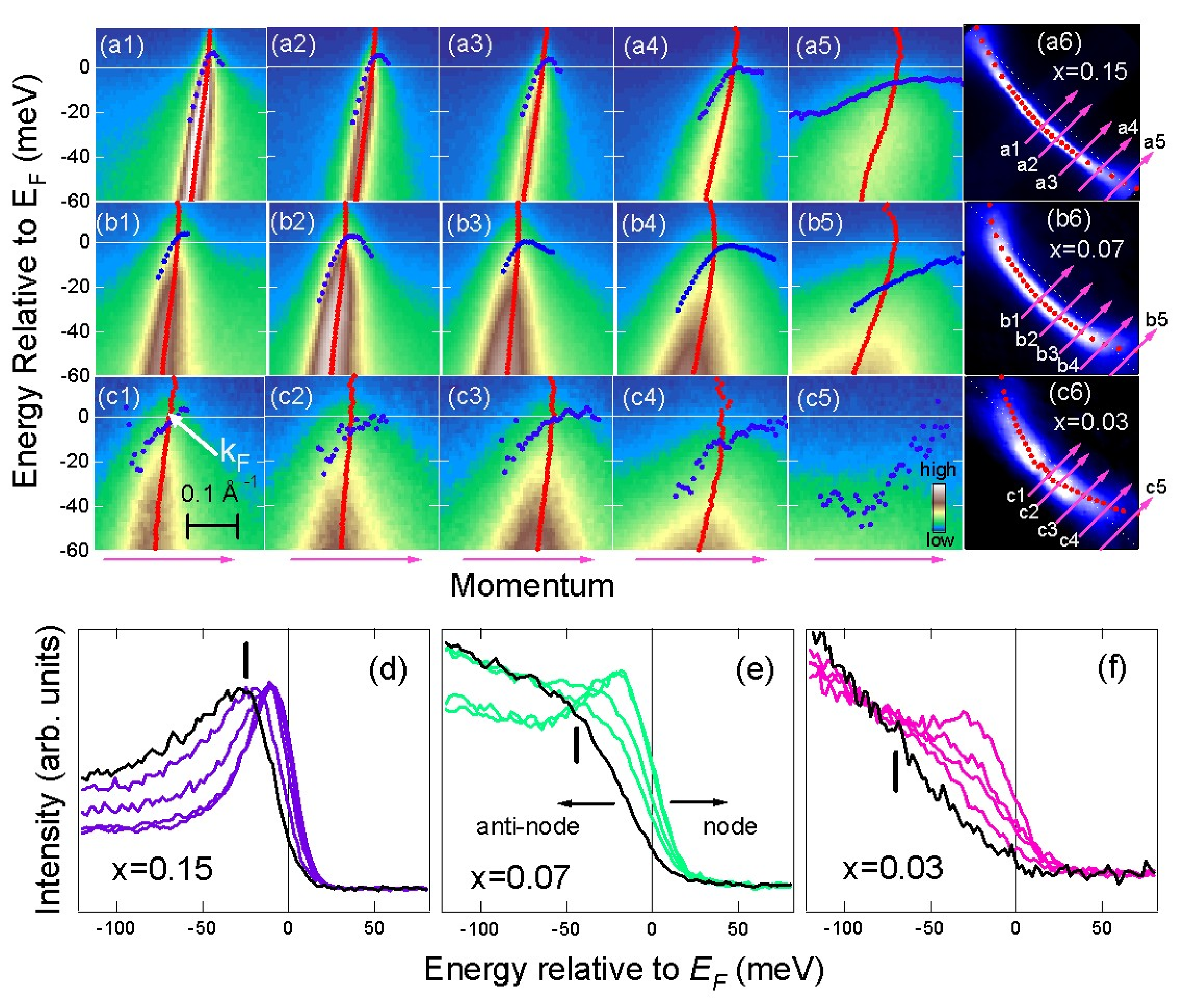}
\caption{\label{EkAll}(Color online) ARPES intensity plot of
La$_{2-x}$Sr$_x$CuO$_4$ (LSCO) for cuts across the Fermi surface.
(a1)-(a5), (b1)-(b5) and (c1)-(c5): Band image plots in
energy-momentum ($E$-$k$) space for $x$=0.15, 0.07 and 0.03,
respectively. Energy dispersions determined by MDC's peaks and
leading edge midpoints (LEM) are shown by red dots and blue dots,
respectively. (a6), (b6) and (c6): Spectral weight mapping at
$E_F$ in momentum space for each doping level. Red dots indicate
Fermi momenta $k_F$ determined by the MDC peak positions at $E_F$.
White dotted lines indicate the antiferromagnetic Brillouin zone
(AFBZ). (d)-(f): EDC's at $k_F$ for each doping level. Black lines
correspond to anti-nodal EDC's and vertical bars represent energy
position of the anti-node gap.}
\end{figure}

High-quality single crystals of LSCO ($x$=0.03, 0.07, 0.15) were
grown by the traveling-solvent floating-zone method. The critical
temperatures ($T_c$'s) of the $x$ = 0.07, 0.15 samples were 14 and
39 K, respectively, and the $x$ = 0.03 samples were
non-superconducting. The ARPES measurements were carried out at
BL10.0.1 of Advanced Light Source (ALS) and at BL-28A of Photon
Factory (PF) using incident photons of linearly polarized 55.5 eV
and circularly polarized 55 eV, respectively. SCIENTA R4000 and
SES-2002 analyzer were used at ALS and PF, respectively, with the
total energy resolution of $\sim$20 meV and momentum resolution of
$\sim$0.02$\pi/a$, where $a$=3.8 \textrm{\AA} is the lattice
constant. The samples were cleaved \textit{in situ} and
measurements were performed from 20 to 155 K. In the measurements
at ALS, the electric field vector $\mathbf{E}$ of the incident
photons lies in the CuO$_2$ plane, rotated by 45 degrees from the
Cu-O bond direction, so that its direction is parallel to the
Fermi surface segment around the nodal region. This geometry
enhances the dipole matrix elements in this $ \mathbf{k}$ region
\cite{yoshidaOD}.

In Fig. \ref{EkAll}, ARPES intensity in energy-momentum space for
various cuts is mapped from the nodal to the anti-nodal
directions. The quasi-particle (QP) band dispersions are
determined by momentum distribution curve (MDC) peak positions and
the Fermi momentum $k_F$ is defined by the momentum where the QP
dispersion crosses the $E_F$. The leading edge midpoints (LEM's)
of the energy distribution curves (EDC's) are plotted by blue dots
around the $k_F$ for each cut. To quantify the energy gap size,
the LEM's at $k_F$ shall be used in the present analysis. As shown
in Fig.\ref{EkAll}(d)-(f), the LEM's at $k_F$ are shifted toward
higher binding energies in going from the node to the anti-node,
indicating an anisotropic gap opening.

\begin{figure}
\includegraphics[width=9cm]{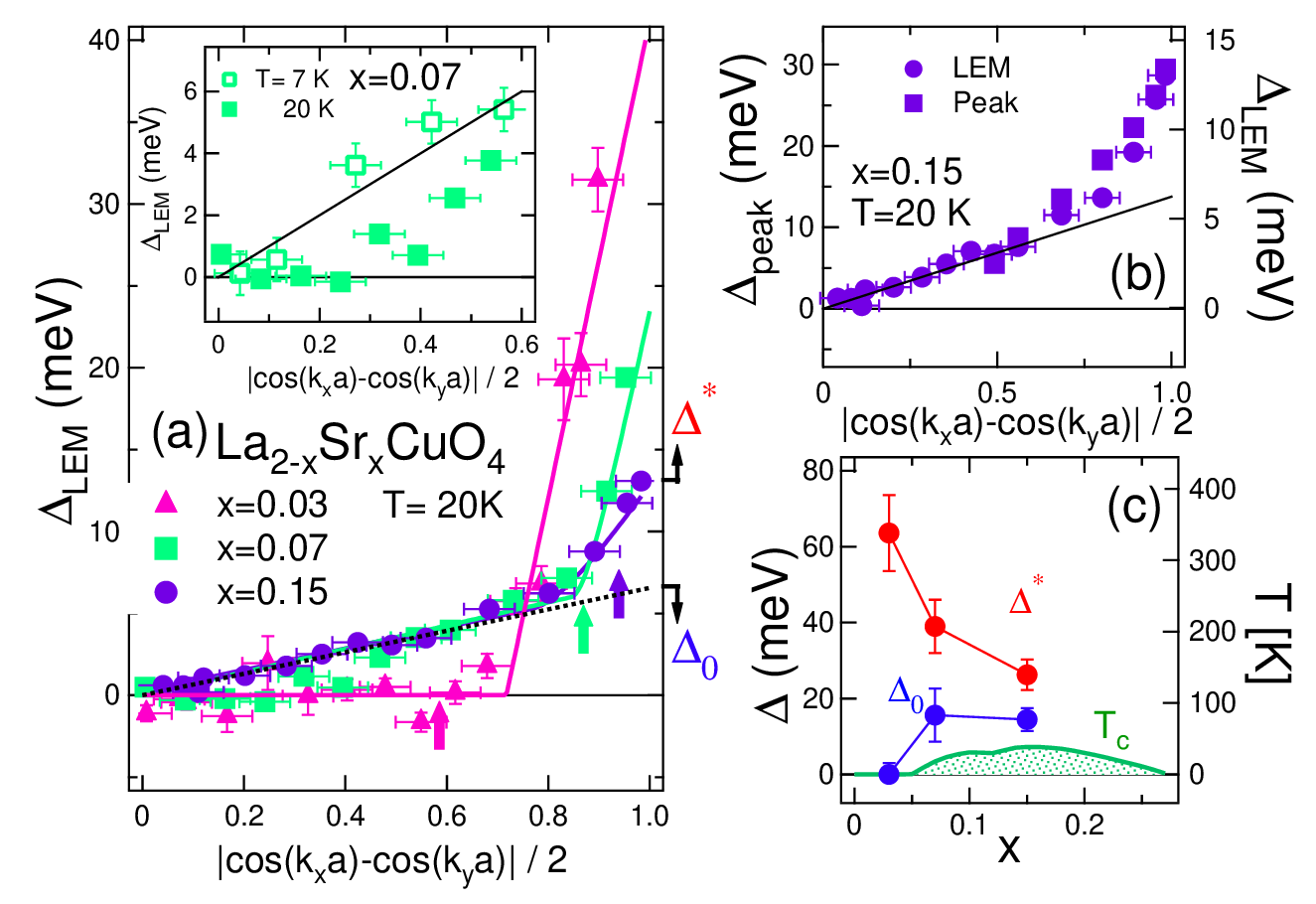}
\caption{\label{LEM_doping}(Color online) Momentum dependence of
the energy gap at $T$= 20 K in LSCO with various doping levels.
(a): Leading edge midpoints (LEM) $\Delta_\mathrm{LEM}$ relative
to that at the node. Vertical arrows represents the boundary of
the AFBZ. Inset shows LEM near the node for $x$=0.07 below and
above $T_c$ (=14 K). (b): Comparison of the peak position
($\Delta_\mathrm{peak}$) and $\Delta_\mathrm{LEM}$ for $x$=0.15,
indicating the relationship $\Delta_\mathrm{peak} \simeq 2.2
\Delta_\mathrm{LEM}$. (c): Doping dependence of $\Delta^*$ and
$\Delta_0$ obtained by assuming the relation in panel (b).}
\end{figure}

The gap sizes have been evaluated from the shift
($\Delta_\mathrm{LEM}$) of the LEM of EDC's relative to the node.
The angular dependence of the gap for each doping is plotted as a
function of the $d$-wave parameter $|\cos(k_xa)-\cos(k_ya)|$/2 in
Fig. \ref{LEM_doping}(a). These plots do not obey the simple
straight line expected for the pure $d$-wave order parameter but
has a kink at $|\cos(k_xa)$-$ \cos(k_ya)|$/2$\cong$0.7-0.9.
Interestingly, the kink occurs near the antiferromagnetic
Brillouin-zone boundary but not exactly on it, as shown by
vertical arrows. Qualitatively the same results have been obtained
for the single-layer cuprates Bi2201 \cite{Kondo} and underdoped
Bi2212 \cite{Tanaka,Lee}. Note that the gaps for $x$=0.07 near the
node are almost closed above $T_c$ (=14 K), but $d$-wave like gap
opens below $T_c$ as shown in the inset.


In order to discuss the character of the energy gaps, we define
two distinct energy scales $\Delta^*$ and $\Delta_0$: $\Delta^*$
from $\Delta_\mathrm{LEM}$ closest to
$|\cos(k_xa)-\cos(k_ya)|$/2=1 and $\Delta_0$ from the
extrapolation of the linear dependence near the node
($|\cos(k_xa)-\cos(k_ya)|/2\sim 0$) toward
$|\cos(k_xa)-\cos(k_ya)|$/2=1, as indicated in panel (a). Since
the $\Delta_\mathrm{LEM}$ is affected by the width of EDC's, the
gap magnitude ($\Delta$) is approximately given by 2-3 times
$\Delta_\mathrm{LEM}$ \cite{Kondo}. As shown in Fig.
\ref{LEM_doping}(b), a relationship
$\Delta=2.2\Delta_\mathrm{LEM}$ well explains both the LEM and
peak shift in the $x$=0.15 data and also explains the data for
$x$=0.03 and 0.07. Therefore, we have assumed this relationship
for analysis of the $\Delta^*$ and $\Delta_0$ as described below.

\begin{figure}
\includegraphics[width=9cm]{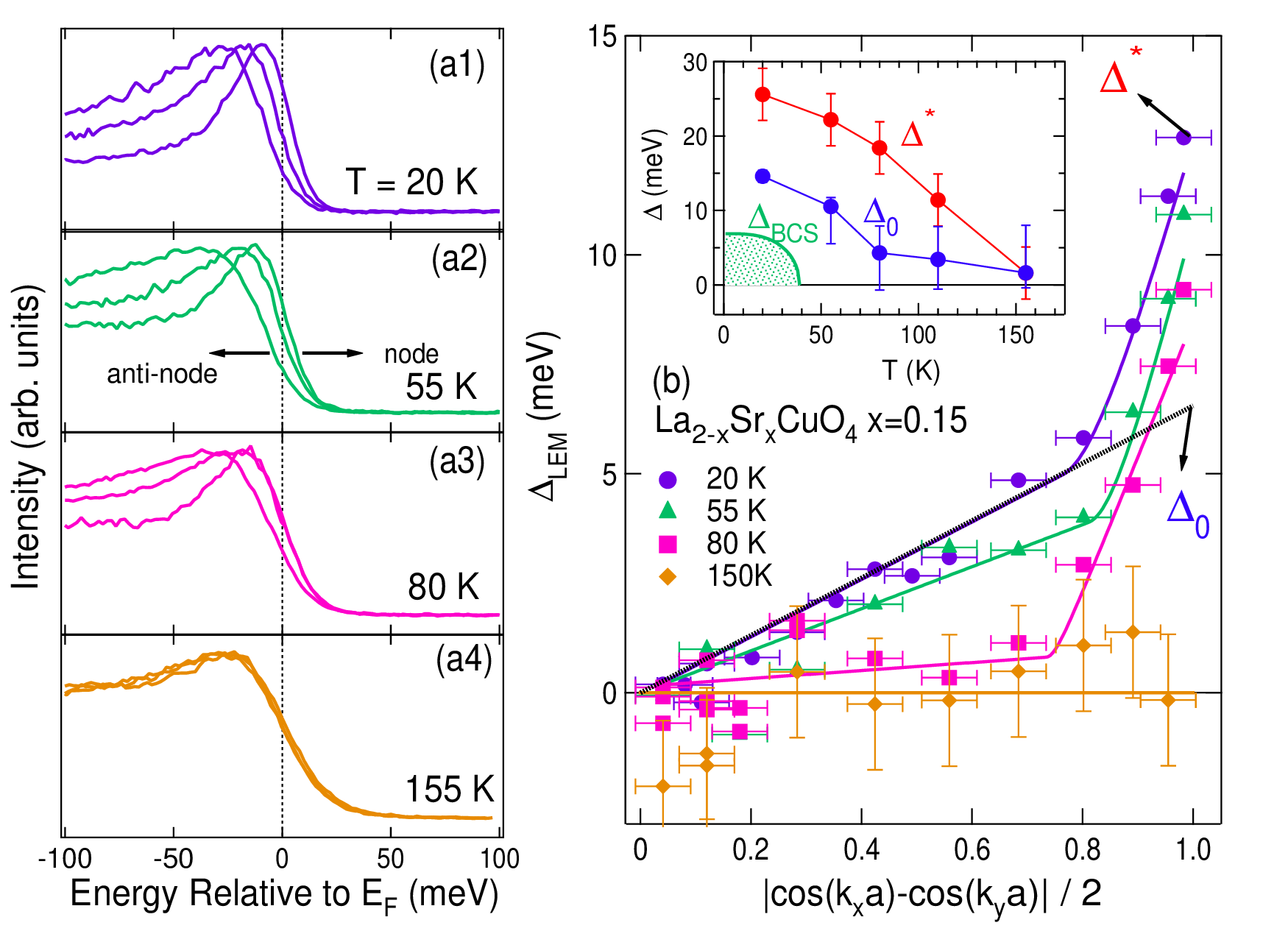}
\caption{\label{LEM_T}(Color online) Momentum dependence of the
energy gap for $x$=0.15 at various temperatures. (a1)-(a4): EDC's
at $k_F$ in the nodal to the anti-nodal directions. (b): Momentum
dependence of $\Delta_\mathrm{LEM}$ along the Fermi surface for
$x$=0.15 at various temperatures. Inset shows the temperature
dependence of $\Delta^*$ and $\Delta_0$ with the assumption
$\Delta=2.2\Delta_\mathrm{LEM}$ as in Fig. \ref{LEM_doping}. The
$d$-wave BCS gap $\Delta_\mathrm{BCS}$(=4.3$k_B T_c/2$) is also
plotted for comparison.}
\end{figure}

In Fig. \ref{LEM_doping}(c), the doping dependence of the observed
$\Delta^*$ and $\Delta_0$ thus deduced are summarized. The doping
dependence of $\Delta^*$ is quantitatively consistent with various
spectroscopic data such as B$_{1g}$-geometry Raman scattering
\cite{Tacon}. $\Delta^*\sim$ 30 meV for the $x$=0.15 sample is
consistent with the previous ARPES results, too \cite{Terashima}.
On the other hand, $\Delta_0$ remains unchanged in going from
$x$=0.15 to $x$=0.07 and vanishes in the non-superconducting
sample $x$=0.03 \cite{Delta0}, similar to the results of the
lightly-doped Bi2212 \cite{Tanaka}. However, vortex-liquid states
suggestive of superconducting states were observed in $x$=0.03
\cite{Li}. The present result $\Delta_0\sim$0 for $x$=0.03 may be
due to the high temperature effects similar to the LEM above $T_c$
near the nodal direction for $x$=0.07.

The temperature dependence of the gap is shown in Fig.
\ref{LEM_T}. In Fig. \ref{LEM_T}(a1)-(a4), EDC's for $x$=0.15
exhibit clear shifts of the LEM between the nodal and anti-nodal
directions in the $T$= 20, 55 and 80 K data. In contrast, the LEM
at $T$=155 K show almost no shift between the nodal and anti-nodal
direction, indicating that the gap is closed on the entire Fermi
surface. The angular dependence of the $\Delta_{LEM}$ for each
temperature are plotted in Fig. \ref{LEM_T}(b). As shown in the
inset, $\Delta^*$ decreases with increasing temperature and closes
at $T^*\sim$150 K, again consistent with $T^*$ obtained from the
angle-integrated photoemission results \cite{Hashimoto}.
$\Delta_0$ also decreases with temperature similar to the decrease
of $T_c$. However $\Delta_0$ seems finite at $T$ =55 K, slightly
above $T_c$. Probably, the gap closes near the node direction
\cite{Meng}, although the low energy scales in LSCO did not allow
us to resolve it.

\begin{figure}
\includegraphics[width=9cm]{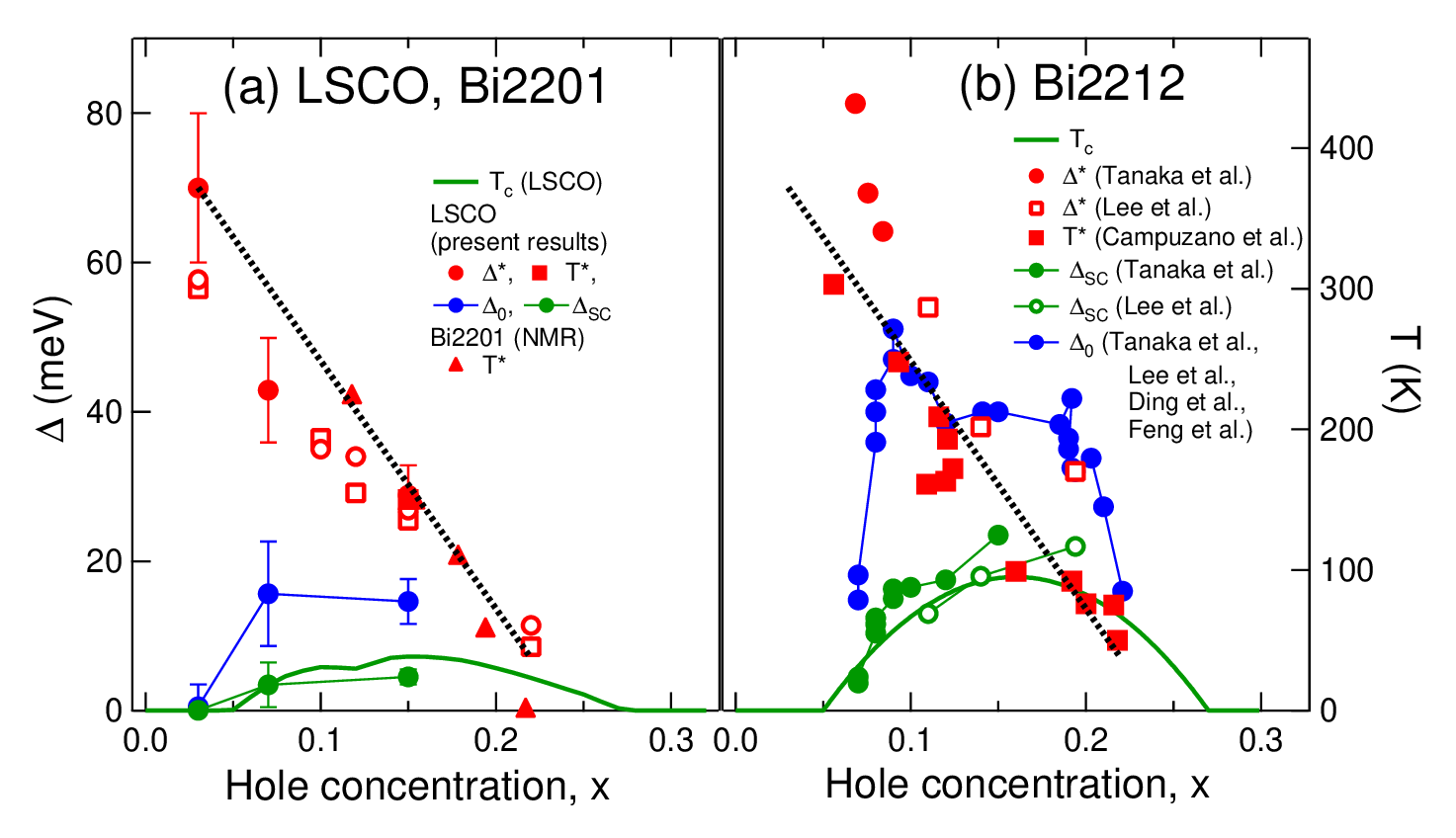}
\caption{\label{Gap_LSCO_Bi2212}(Color online) Doping dependence
of the characteristic energies ($\Delta^*, \Delta_0$) and
temperatures ($T^*, T_c$) for single-layer cuprates (LSCO, Bi2201)
(a) and double-layer cuprates Bi2212 (b). Gap energies $\Delta$
and temperatures $T$ have been scaled as $2\Delta=4.3k_BT$ in both
panels. Parameter values have been taken from NMR results for
Bi2201 \cite{Zheng}, and ARPES for Bi2212
\cite{Tanaka,Lee,Campuzano,Ding,Feng}. }
\end{figure}

Now, let us compare the two gap energy scales of LSCO with those
of other high-$T_c$ cuprates to clarify their relation to $T_c$.
In Fig. \ref{Gap_LSCO_Bi2212}(a), the doping dependences of
$\Delta^*, \Delta_0$ and $T^*$ for LSCO and another single-layer
cuprate Bi2201 are plotted. In the same manner, those for
double-layer Bi2212 which has about twice higher $T_c$ than those
of LSCO are plotted in Fig. \ref{Gap_LSCO_Bi2212}(b).
Interestingly, the doping dependences of $\Delta^*$ of all these
samples approximately scale with $T^*$ following the relationship
$2\Delta^*/k_BT^*=4.3$, reminiscent of the $d$-wave BCS
relationship \cite{Maki}. Furthermore, these data fall on
approximately the same lines for all the compounds irrespective of
the different $T_c$ as indicated in Fig \ref{Gap_LSCO_Bi2212}(a)
and (b). Especially, pseudogap temperatures for optimally doped
Bi$_2$Sr$_2$Ca$_2$Cu$_3$O$_{10}$ (Bi2223) \cite{Sato} and Bi2201
\cite{Kondo} are both $T^*\sim 150K$, similar to the present
result of LSCO $T^*\sim 140K$, although they have very different
$T_c$'s. Therefore, we speculate that $\Delta^*$ is an universal
property of a single CuO$_2$ plane and is not much affected by its
chemical environment \cite{disorder}. One possible explanation for
the material independence of $\Delta^*$ is that its magnitude is
determined by $J$, since the exchange interaction $J$ is almost
material independent. A pseudogap originated from
antiferromagnetic spin fluctuations \cite{Prelovsek} or RVB-type
spin singlet formation \cite{fukuyama} has its origin in $J$.

In contrast to $\Delta^*$, the $d$-wave order parameter $\Delta_0$
of Bi2212 is twice as large as those of LSCO, reminiscent of the
difference in the magnitude of $T_c$. The strong material
dependences of $\Delta_0$ mean that $\Delta_0$ is not a property
only of a single CuO$_2$ plane but also influenced by the
environment such as the apical oxygens or the block layers and/or
the neighboring CuO$_2$ planes in multilayer cuprates. Namely, the
number of CuO$_2$ layers and the distance of the apical oxygen
atoms (in block layers) from the CuO$_2$ plane are important
factors for the superconducting gap and hence $T_c$. Within the
model Hamiltonian description of the high-$T_c$ cuprates, the
effect from outside the CuO$_2$ plane has been modelled using the
distant-neighbor hopping parameters $t^\prime$ and
$t^{\prime\prime}$ \cite{Pavarini}, which are affected by the
$p_z$ orbital of the apical oxygen and the position of the empty
Cu 4$s$ orbital and characterize the details of band dispersions.
In other words, the ($\pi$,0) pseudogap does not depend on details
of the band structure nor on the parameters $t^\prime$,
$t^{\prime\prime}$, but only on $t$ and/or $J$.

If the pseudogap in the anti-nodal region precludes contribution
to the superconductivity and the superconductivity comes mainly
from the near-nodal region in the underdoped cuprates, the
``effective" superconducting gap $\Delta_\mathrm{sc}\propto$
(Fermi arc length) $\times \Delta_0$ rather than $\Delta_0$ would
be more directly related to $T_c$ \cite{Oda}. Here, the Fermi arc
is defined by the momentum region where the energy gap closes just
above $T_c$. According to the high-resolution ARPES, the arc
length for LSCO ($x$=0.15) is $\sim$ 30 \% of the entire Fermi
surface \cite{Terashima}, which is consistent with the present
results with $T$=80 K. For $x$=0.07, the arc length is $\sim$ 20
\% as seen in the inset of Fig. \ref{LEM_doping}. Using these
value for LSCO, the doping dependence of $\Delta_\mathrm{sc}$=
(arc length) $\times \Delta_0$ is plotted in Fig.
\ref{Gap_LSCO_Bi2212} (a). In the same manner,
$\Delta_\mathrm{sc}$ for Bi2212 were determined by using the arc
length reported in Ref. \cite{Lee} [Fig.
\ref{Gap_LSCO_Bi2212}(b)]. The plotted $\Delta_\mathrm{sc}$
approximately agree with the dome of $T_c$ through the BCS formula
2$\Delta_{sc}$=4.3$k_B T_c$. Particularly, the decrease of $T_c$
with underdoping can be ascribed to the reduction of the Fermi arc
length together with the $\Delta_0$, which remains nearly constant
till $x\sim$0.07 and then drops. As for the non-superconducting
$x$=0.03 sample, the arc length may be too short to produce
sufficient carriers for superconductivity or the nodal spectra may
have a small gap due to localization as seen in the transport
properties \cite{Ando}.

In summary, we have performed an ARPES study of LSCO to
investigate the momentum, doping and temperature dependences of
the energy gap from the lightly-doped to optimally doped regions.
We have clearly shown a signature of the two distinct energy gap
scales, $\Delta^*$ and $\Delta_0$. From comparison of the present
results with those of other cuprates, we have found that the
magnitude of $\Delta^*$ is almost material-independent, suggesting
that the pseudogap is a distinct phenomenon from
superconductivity. On the other hand, $\Delta_0$ exhibits a large
difference between materials, reflecting the different
superconducting properties including the different $T_c$'s. Using
the obtained two-gap parameters in conjunction with the Fermi arc
picture \cite{Oda}, we have obtained the magnitude of the
``effective" superconducting gap in the underdoped region and
consistently explained the doping dependence of $T_c$ in LSCO as
well as in Bi2212. The present results enforce the picture of
superconductivity on the Fermi arc and clarify how $T_c$
disappears in the underdoped region. Since the observed material
dependence of $\Delta_0$ is a crucial factor for the high-$T_c$
superconductivity, the relationship between $\Delta_0$ and other
model parameters such as $t^\prime$ and $t^{\prime\prime}$, the
number of CuO$_2$ planes, the apical oxygen - Cu distance, and
possibly electron-phonon coupling, has to be clarified in future
studies.

We are grateful to C. M. Ho, M. Ido, G.-q. Zheng and C.
Panagopoulos for enlightening discussions. This work was supported
by a Grant-in-Aid for Scientific Research in Priority Area
``Invention of Anomalous Quantum Materials", Grant-in-Aid for
Young Scientists from the Ministry of Education, Science, Culture,
Sports and Technology and the USDOE contract DE-FG03-01ER45876 and
DE-AC03-76SF00098. Y.A. was supported by KAKENHI 19674002 and
20030004. ALS is operated by the Department of Energy's Office of
Basic Energy Science, Division of Materials Science. Experiment at
Photon Factory was approved by the Photon Factory Program Advisory
Committee (Proposal No. 2006S2-001).

\bibliography{TwoGap}

\begin{thebibliography}{10}
\bibitem{Delta0}
For $x$=0.03, although $\Delta_\mathrm{LEM}\sim$0 on most of the Fermi surface,
  spectral weight is appreciable only near the node, yielding the short Fermi
  arc.
\end{thebibliography}

\end{document}